\pdfoutput=1
\documentclass[a4paper,aip,american,citeautoscript,floatfix,jcp,pdftex,showpacs,superscriptaddress,twopage%
longbibliography,%
reprint,final,twocolumn%
]{revtex4-1}
\usepackage{amsmath,amssymb}
\usepackage[main=american,german]{babel}
\usepackage{graphicx}
\usepackage[T1]{fontenc}
\usepackage[utf8]{inputenc}
\usepackage{microtype}
\usepackage{subfigure}
\usepackage{textcomp}
\usepackage[textsize=footnotesize]{todonotes}
\usepackage{xspace}
\usepackage{hyperref, hypernat}
\usepackage[todos]{CMImacros}

\newlength{\figwidth}
\newcommand{\cfeldesy}{\affiliation{Center for Free-Electron Laser Science,
      Deutsches Elektronen-Synchrotron DESY, Notkestrasse 85, 22607 Hamburg, Germany}}%
\newcommand{\uhhcui}{\affiliation{The Hamburg Center for Ultrafast Imaging,
      University of Hamburg, Luruper Chaussee 149, 22761 Hamburg, Germany}}%
\newcommand{\uhhphys}{\affiliation{Department of Physics, University of Hamburg,
      Luruper Chaussee 149, 22761 Hamburg, Germany}}%

\begin{document}
\title{Improved spatial separation of neutral molecules}%
\author{Jens S.\ Kienitz}\cfeldesy\uhhcui%
\author{Karol Długołęcki}\cfeldesy%
\author{Sebastian Trippel}\email{sebastian.trippel@cfel.de}%
\cfeldesy\uhhcui%
\author{Jochen Küpper}\email{jochen.kuepper@cfel.de}%
\homepage{https://www.controlled-molecule-imaging.org}%
\cfeldesy\uhhcui\uhhphys%
\date{\today}%
\begin{abstract}\noindent%
   We have developed and experimentally demonstrated an improved electrostatic deflector for the
   spatial separation of molecules according to their dipole-moment-to-mass ratio. The device
   features a very open structure that allows for significantly stronger electric fields as well as
   for stronger deflection without molecules crashing into the device itself. We have demonstrated
   its performance using the prototypical OCS molecule and we discuss opportunities regarding
   improved quantum-state-selectivity for complex molecules and the deflection of unpolar molecules.
\end{abstract}
\maketitle%
\noindent%

\section{Introduction}
\label{sec:introduction}
The history of the deflection of neutral atoms and molecules in vacuum began in 1922 with the famous
Stern-Gerlach-experiment~\cite{Stern:ZP7:249, Gerlach:ZP9:349, Rabi:PR55:526}, which exploited the
Zeeman effect of silver atoms and the corresponding force in an inhomogeneous magnetic field. Just a
few years later the deflection of molecules by an inhomogeneous electric field was
demonstrated~\cite{Kallmann:ZP6:352, Wrede:ZP44:261}. Various deflector shapes and techniques have
evolved to optimize the process~\cite{Chamberlain:PR129:677, Filsinger:JCP131:064309,
   DeNijs:PCCP13:19052, Chang:IRPC34:557} and an overview of various early experimentally employed
deflection-field geometries was already presented in Ramsey's textbook on molecular
beams~\cite{Ramsey:MolBeam:1956}. Modern experiments typically employed geometries approximating
two-wire fields~\cite{Chamberlain:PR129:677, Filsinger:JCP131:064309}. The technique allowed for a
wide range of application, such as the separation of molecules in individual rotational
states~\cite{Holmegaard:PRL102:023001, Wohlfart:PRA77:031404, Putzke:PCCP13:18962}, of
conformers~\cite{Filsinger:ACIE48:6900, Filsinger:PRL100:133003, Kierspel:CPL591:130}, and of
specific molecular clusters~\cite{Trippel:PRA86:033202}. Additionally, the deflector enables the
separation of polar molecules from the seed gas, which is relevant, for instance, for gas-phase
diffraction experiments~\cite{Kuepper:PRL112:083002, Filsinger:PCCP13:2076, Barty:ARPC64:415,
   Mueller:thesis:2016}.

Alternatively, eigenstates of small polar neutral molecules can be separated using switched electric
or magnetic fields~\cite{Meerakker:NatPhys4:595, Schnell:ACIE48:6010, Bell:MP107:99,
   Hogan:PCCP13:18705, Meerakker:CR112:4828, Bethlem:JPB39:R263, Wohlfart:PRA77:031404,
   Putzke:JCP137:104310}. Large molecular ions were separated according to their shape using
ion-mobility techniques~\cite{Bohrer:ARAC1:293, Wyttenbach:ANPC65:175}. Furthermore, laser fields
have been used to deflect and decelerate neutral molecules~\cite{Stapelfeldt:PRL79:2787,
   Zhao:PRL85:2705, Fulton:PRL93:243004, Sun:PRL115:223001}, which also works for non-polar
molecules. Laser-based electric deflection has demonstrated
state-separability~\cite{Sun:PRL115:223001}, but is is limited by very small volumes, typically only
manipulating a small subset of the molecular beam. The polarizability interaction was also used in
the deceleration of hydrogen molecules in Rydberg states~\cite{Procter:CPL374:667} and in
weak-deflection polarizability measurements~\cite{Scheffers:PhysZ35:16, Chamberlain:PR129:677}. In
addition inhomogeneous ac- and dc-electric fields are predicted to manipulate the average deflection
angle and its distribution when combined with strongly aligned molecular
samples~\cite{Gershnabel:PRL104:153001, Gershnabel:JCP135:084307}.

We have developed a Stark deflector for increased separation of cold molecular species in a beam
that utilizes an improved geometry and supports stronger fields. The geometry is similar to
previously proposed~\cite{Stefanov:MeasSciTech19:055801, DeNijs:PCCP13:19052} and
utilized~\cite{Eibenberger:NJP13:043033} deflectors, but is unique in it's applicability for the
separation of molecular species in dense molecular beams~\cite{Chang:IRPC34:557}. Here, we
experimentally demonstrate its performance for the separation of the eigenstates of carbonyl sulfide
(OCS). We have characterized the deflection power through column-density measurements and the
separation of eigenstates through state-specific mixed-field orientation
measurements~\cite{Nielsen:PRL108:193001, Kienitz:CPC17:3740}. Moreover, we have performed
simulations on the deflection of non-polar molecules and discuss the feasibility of such
experiments.

\section{Theory}
\label{sec:theory}
\subsection{Stark effect}
\label{sec:stark-effect}
In first order, the interaction of a molecule with a dc electric field, the Stark effect, is
governed by the so-called permanent dipole moment $\boldsymbol{\mu}$~\cite{Gordy:MWMolSpec,
  Chang:IRPC34:557}:
\begin{equation}
   W^{(1)} = -\boldsymbol{\mu}\cdot\boldsymbol{\epsilon} = -\mu\epsilon\costheta
   \label{eq:stark:dip}
\end{equation}
and by the polarizability tensor $\boldsymbol{\alpha}$, or induced-dipole $\boldsymbol{\mu_\text{ind}}$,
interaction~\cite{Gordy:MWMolSpec}:
\begin{equation}
   W^{(2)} = \frac{1}{2}\boldsymbol{\epsilon}\boldsymbol{\alpha}\boldsymbol{\epsilon}
   = -\boldsymbol{\mu_\text{ind}}\cdot\boldsymbol{\epsilon}
   \label{eq:stark:pol}
\end{equation}
The effective, space-fixed, dipole moment $\mueff=-\partial{W}/\partial\epsilon$, the derivative
of the energy $W=W^{(1)}+W^{(2)}$ with respect to the electric field, directly describes the
interaction strength of the molecule with the field and the corresponding force in an inhomogeneous
electric field~\cite{Bethlem:JPB39:R263, Chang:IRPC34:557}:
\begin{align}
   W &= \mueff\epsilon \\
   \boldsymbol{F} &= \mueff\boldsymbol{\nabla}\epsilon
   \label{eq:force}
\end{align}

\begin{figure}
   \includegraphics[width=\figwidth]{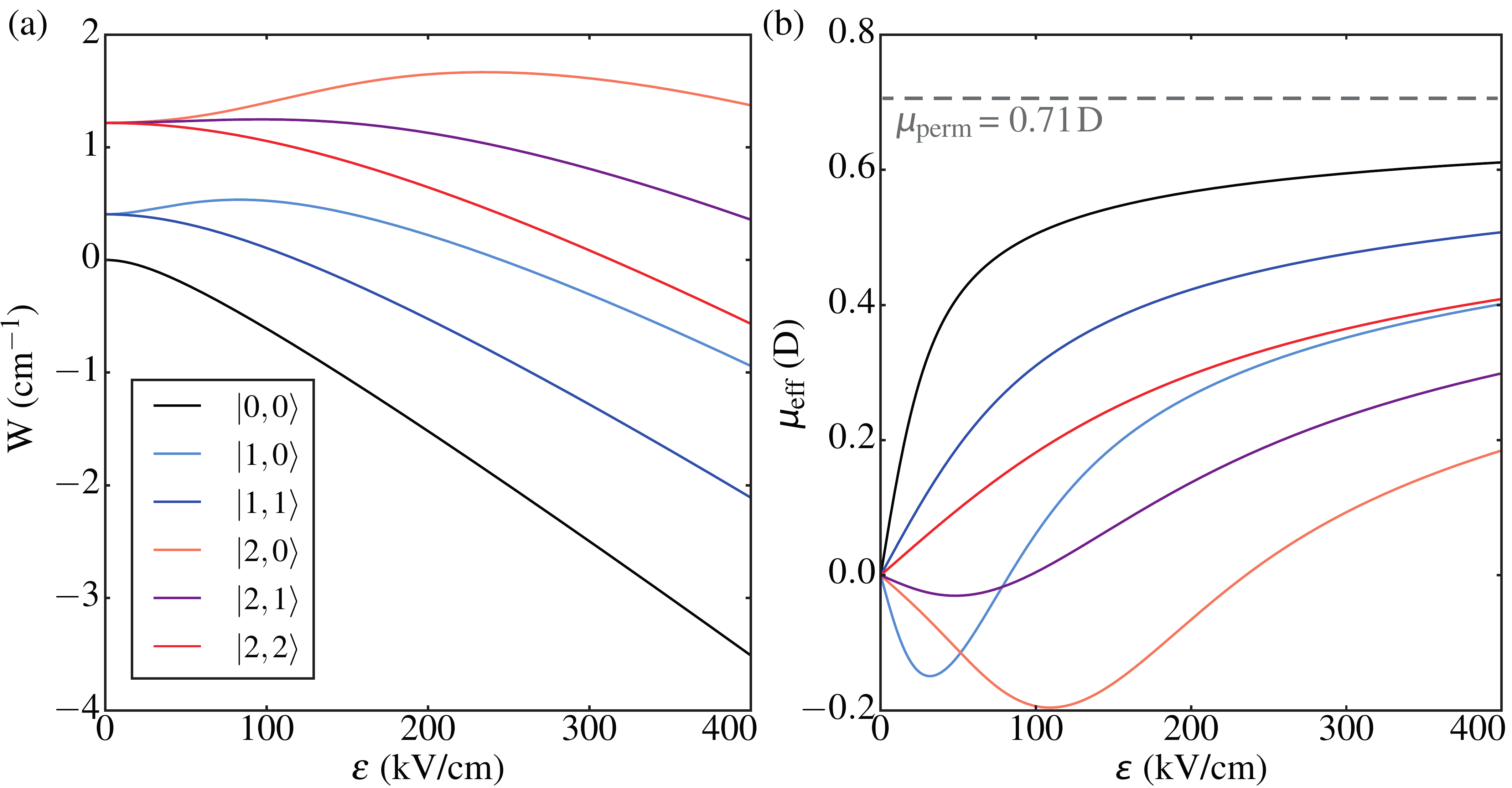}
   \caption{(Color online) (a) Energies of the lowest-energy states of OCS in a static electric
      field and (b) the corresponding effective dipole moments. The gray line depicts the
      strong-field limit of the permanent dipole moment.}
   \label{fig:ocs-stark}
\end{figure}
Stark effect calculations were performed using a modified version of the \textsc{CMIstark} program
package~\cite{Chang:CPC185:339} with added polarizability matrix elements according to
\eqref{eq:stark:pol} \cite{Gordy:MWMolSpec}. The resulting energies and effective dipole moments of
the lowest-energy states of OCS are shown in \autoref{fig:ocs-stark}. For these field strengths the
effect of the polarizability is below 1~\% and is not visible in the data in
\autoref{fig:ocs-stark}.

\subsection{Deflector design}
\label{sec:deflector_design}
The design of the deflector was governed by the need to accept dense cold molecular beams, such as
generated in high-pressure expansions from Even-Lavie valves~\cite{Even:AIC2014:636042,
  Chang:IRPC34:557}. These beams have internal temperatures below 1~K, but in order to avoid beam
heating require large, multi-millimeter diameter skimmers and correspondingly large mechanical
apertures for the deflector. Furthermore, strong electric fields and field gradients allow for
strong deflection and separation. The fields should be strong enough for the molecules to be in the
pendular regime~\cite{Bethlem:JPB39:R263, Chang:IRPC34:557} and the field gradient should be fairly
constant over the cross-section covered by the molecular beam. In high-voltage breakdown tests we
were able to generate static electric fields up to 750~kV/cm before breakdowns occurred, however,
these breakdown voltages do also depend on the actual vacuum
gap~\cite{Latham:HighVoltageVaccumInsulation} and, thus, ask for small structures, while constant
gradients again ask for large mechanical structures.

Here, we settled on a mechanical aperture of the deflector larger than 1.5~mm with a 1.5~mm-diameter
aperture conical skimmer before the device far enough from the valve to avoid choking and
heating. We decided to limit applied voltages to $\pm30$~kV to allow for corresponding feedthroughs.

\begin{figure}
   \includegraphics[width=\figwidth]{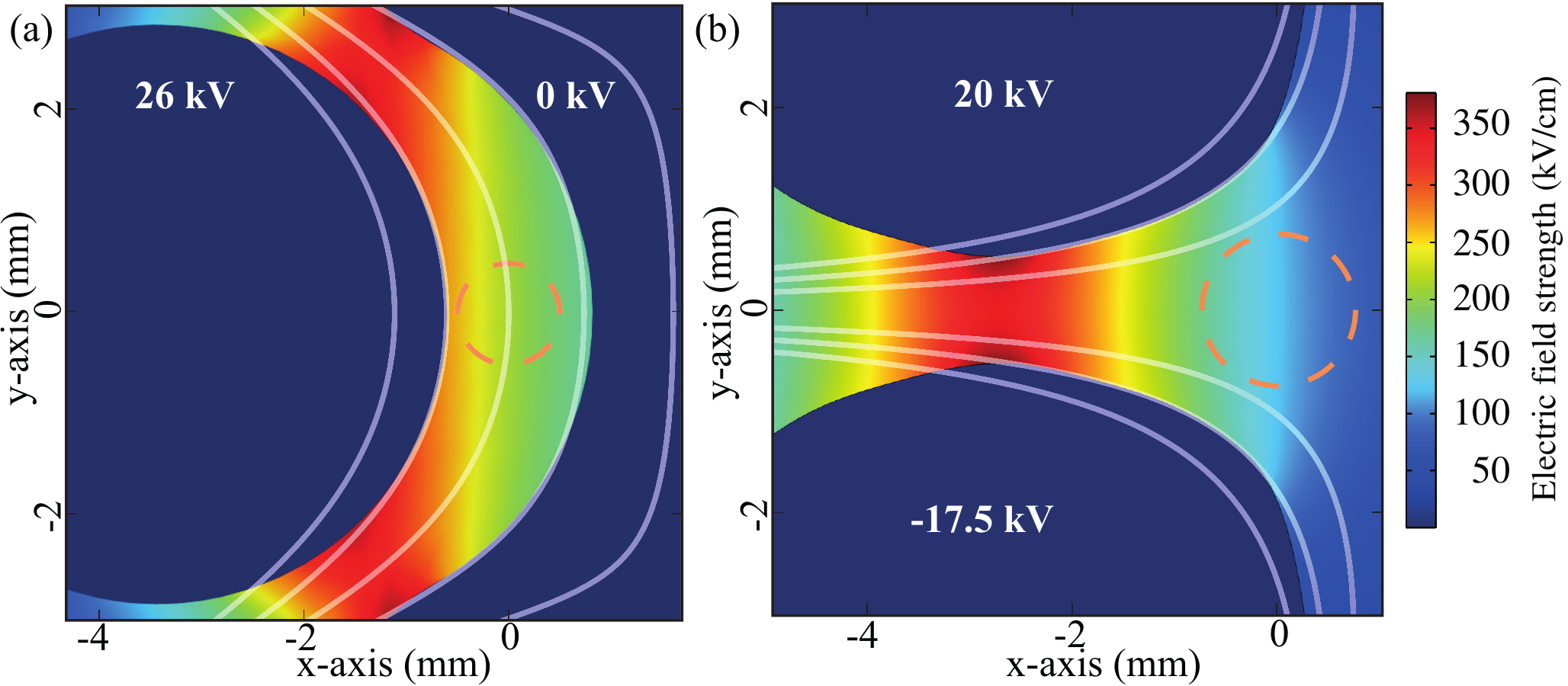}
   \caption{(Color online) Shape and electric fields of the (left) $a$-type and (right) $b$-type
      deflector geometries. The numerically calculated electric field strengths for the actual
      electrode geometries are depicted in color coding and white lines depict the equipotential
      lines according to \eqref{eq:potential-multipole}, the $a_i/b_i$ coefficients specified in
      \autoref{tab:parameter}, and for $\phi/\phi_0$ ratios of (a) $-20,-10,0,10,20$ (left to right)
      and (b) of $0.6, 1.0,\text{~and~}1.4$ (inside to outside). The dashed orange circles represent
      the utilized skimmers and marks the position and size of the incoming molecular beam. In both
      cases molecules in high-field-seeking states are deflected to the left; in the experiments
      discussed below the deflector is rotated counterclockwise by \degree{90}.}
   \label{fig:shape}
\end{figure}

The experimental realization of fields that satisfy such conditions was discussed
before~\cite{Ramsey:MolBeam:1956, Auerbach:JCP45:2160, Bethlem:JPB39:R263, DeNijs:PCCP13:19052}. We
express the shape of the deflector by a equipotential line of a two-dimensional multipole expansion
of the electric potential~\cite{Kalnins:RSI73:2557, Bethlem:JPB39:R263, DeNijs:PCCP13:19052}:
\begin{align}
  \Phi(x,y) = \Phi_0 \Bigg[& \sum_{n=1}^{\infty} \frac{a_n}{n} \left(\frac{r}{r_0}\right)^n
                             \cos\left(n \theta\right) \notag\\
                           & + \sum_{n=1}^{\infty} \frac{b_n}{n} \left(\frac{r}{r_0}\right)^n
                             \sin\left(n \theta\right) \Bigg]
                             \label{eq:potential-multipole}
\end{align}
with $r=\sqrt{x^2+y^2}$ and $\theta=\tan^{-1}(y/x)$; $r_0$ and $\Phi_0$ are scaling factors of the
geometric size and the potential values, respectively. To obtain deflection in $x$ direction either
$a_n$ or $b_n$ can be set to zero, which results in a so called ``$a$-type'' and ``$b$-type''
deflector according to the non-zero coefficients, respectively. The influence of these coefficients
were discussed extensively elsewhere~\cite{Auerbach:JCP45:2160, Bethlem:JPB39:R263,
   DeNijs:PCCP13:19052}. We use the $b$-type shape to allow for an open mechanical structure along
the deflection direction. Moreover, we allow for arbitrarily complex numerically defined electrode
shapes to most closely follow the equipotential lines. The parameters obtained from a numerical
optimization of the $b_n$ ($n=1,2,3$) coefficients~\cite{Kienitz:thesis:2016} as well as the
parameters of previously used $a$-type deflector are specified in \autoref{tab:parameter}.
\begin{table}
   \centering
   \begin{tabular}{lcc}
     \hline\hline
     & ``$a$-type'' & ``$b$-type'' \\
     \hline
     $a_1$/$b_1$ & 0.50 & 1.94 \\
     $a_2$/$b_2$ & 0.49 & -4.80 \\
     $a_3$/$b_3$ & 0.42 & 1.00 \\
     \hline\hline
   \end{tabular}
   \caption{Parameters according to \eqref{eq:potential-multipole} for the two
      deflector geometries discussed in the text. The mechanical scale in our simulations was
      fixed at $r_0= 3.3$~mm.}
   \label{tab:parameter}
\end{table}
Cross sections of the corresponding fields, equipotential lines, and electrode geometries are
depicted in \autoref{fig:shape}. The applied voltages correspond to a maximum field strength of
363~kV/cm, which corresponds to the experiments discussed below.

Actual three-dimensional structures were obtained by extruding the calculated 2D cross sections and
rounding the ends with a radius of 2~mm. For the $b$-type deflector the equipotential lines were
reproduced for $\phi=1$ and $x>-2.73$~mm and mirrored to $x<-2.73$~mm. The resulting cusp was
removed by approximating this geometry by a tenth-order even polynomial Taylor expansion over the
range $-5.84~\text{mm}<x<0.37~\text{mm}$. This symmetry of the deflector allows to deflect
molecules in both direction depending on the relative positioning of deflector and beam.

\section{Experimental setup}
\label{sec:setup}
\begin{figure}
   \centering
   \includegraphics[width=\figwidth]{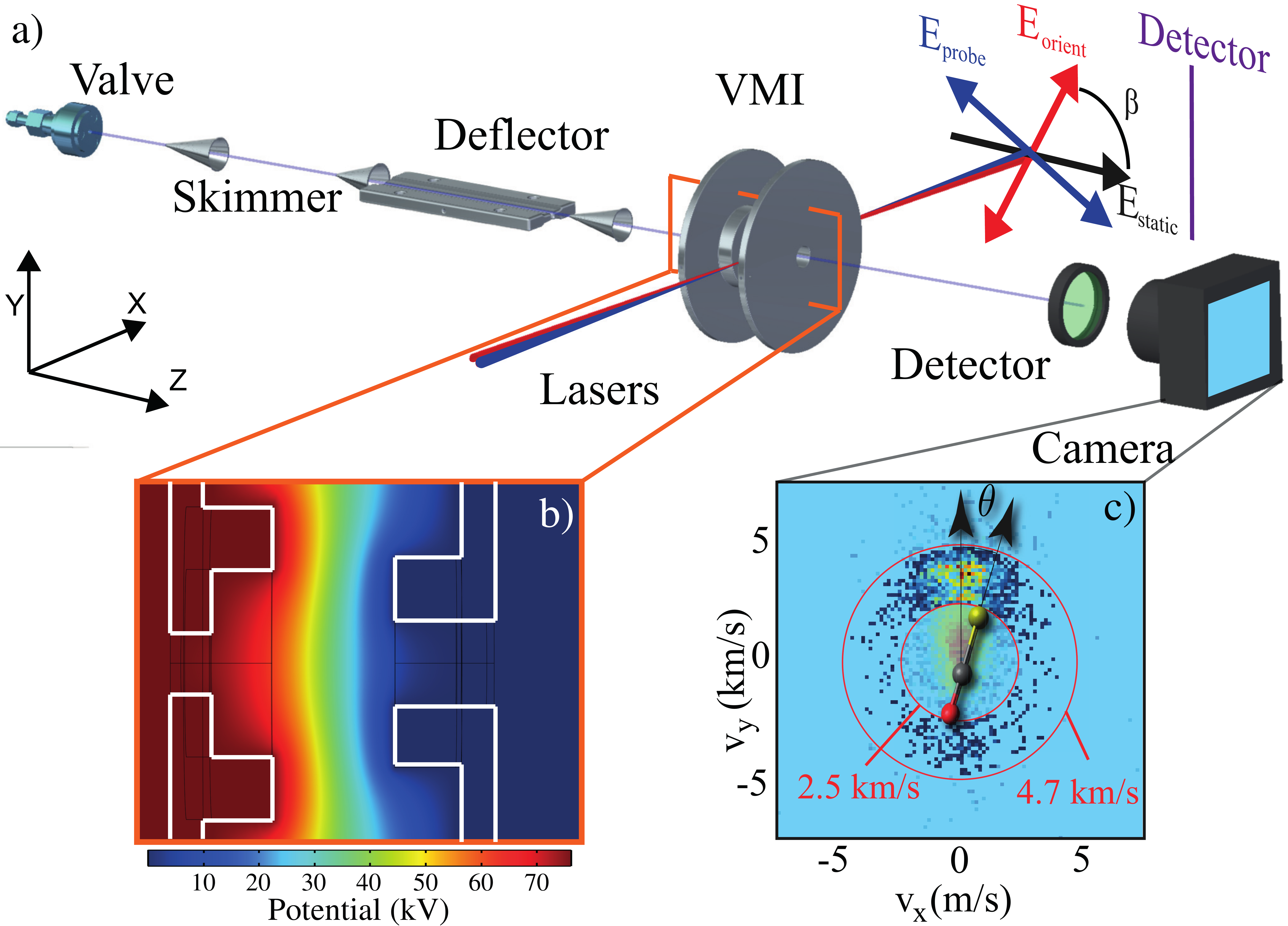}
   \caption{(Color online) (a) Experimental setup with the definition of the axis system. The angle
      between the static electric field and the polarization vector of the control laser pulse is
      defined by $\beta$. (b) Electric field in the velocity-map imaging spectrometer, white lines
      depict the electrode surfaces. (c) Velocity-map image of S$^+$ ions from OCS with red circles
      depicting the velocities defining the region of interest for the data analysis. See text for
      details.}
   \label{fig:experimental_setup_btd}
\end{figure}

The experimental setup is sketched in \autoref{fig:experimental_setup_btd}.
A mixture of 500~ppm OCS seeded in 100~bar of helium was expanded into vacuum through an
Even-Lavie-valve~\cite{Even:JCP112:8068} at a repetition rate of 250~Hz. Two conical skimmers
($\varnothing4$~mm and $\varnothing1.5$~mm) were placed 10.7~cm and 21.6~cm downstream from the
nozzle, respectively. The Stark deflector was located 4.4~cm behind the tip of the second skimmer.
The deflector was wire-eroded and electro-polished. After careful high-voltage conditioning of the
electrodes a voltage difference of 38~kV was applied, which corresponds to the fields shown in
\autoref{fig:shape}b. A third, transversely adjustable skimmer ($\varnothing1.5$~mm) was placed
4.7~cm downstream of the end of the deflector.

Two laser pulses, both provided by amplified femtosecond laser system with a central wavelength of
800~nm and propagating along a common line, were used to orient and probe the
molecules~\cite{Trippel:MP111:1738}. The temporal profile of the control laser pulse had a sawtooth
shape with a slow rising edge (680~ps, 2.5-97.5\%) and a fast falling edge (190~ps). The spatial
intensity profile in the interaction volume had widths of $\sigma=16~\um$ and $\sigma'=21~\um$ along
the two principal axes of the profile. The principal axis were rotated by \degree{20} from the $Y$
axis. The peak intensity of the control laser was $\Icontrol\approx5\times10^{11}~\text{W/cm}^2$.
Short pulses of 30~fs laser focused to $\sigma_1=\sigma_2=20~\um$ and peak intensities of
$\Iprobe\approx10^{14}~\text{W/cm}^2$ were used to multiple ionize the molecules to induce Coulomb
explosion. The relative delay between both laser pulses was controlled with a motorized delay stage.
Both lasers pulses were linearly polarized perpendicular with respect to each other. The control
laser polarization is at a fixed angle $\beta=\degree{45}$ with respect to the $Z$-axis.

A high-voltage velocity-map-imaging spectrometer was used to record the momenta of S$^+$ ions as a
signature of the molecules' directions at the time of ionization~\cite{Kienitz:CPC17:3740}.
Vertical molecular beam profiles, with and without voltages applied to the deflector, were recorded
by moving the skimmer and the laser focus through the molecular beam and recording the integrated
number of S$^+$ ions at each position. The envelope of all events over all skimmer positions yields
the overall beam profile. Furthermore, we derived the degree of orientation across the molecular
beam with the control laser pulses applied from the angular distribution of S$^+$ ions in the region
of interest ($2500~\text{m/s}<v_{xy}<4700~\text{m/s}$) shown in
\autoref{fig:experimental_setup_btd}.

\section{Results}
\label{sec:results}
\subsection{Deflection of OCS}
\label{sec:deflection-ocs}
\begin{figure}
   \centering
   \includegraphics[width=\figwidth]{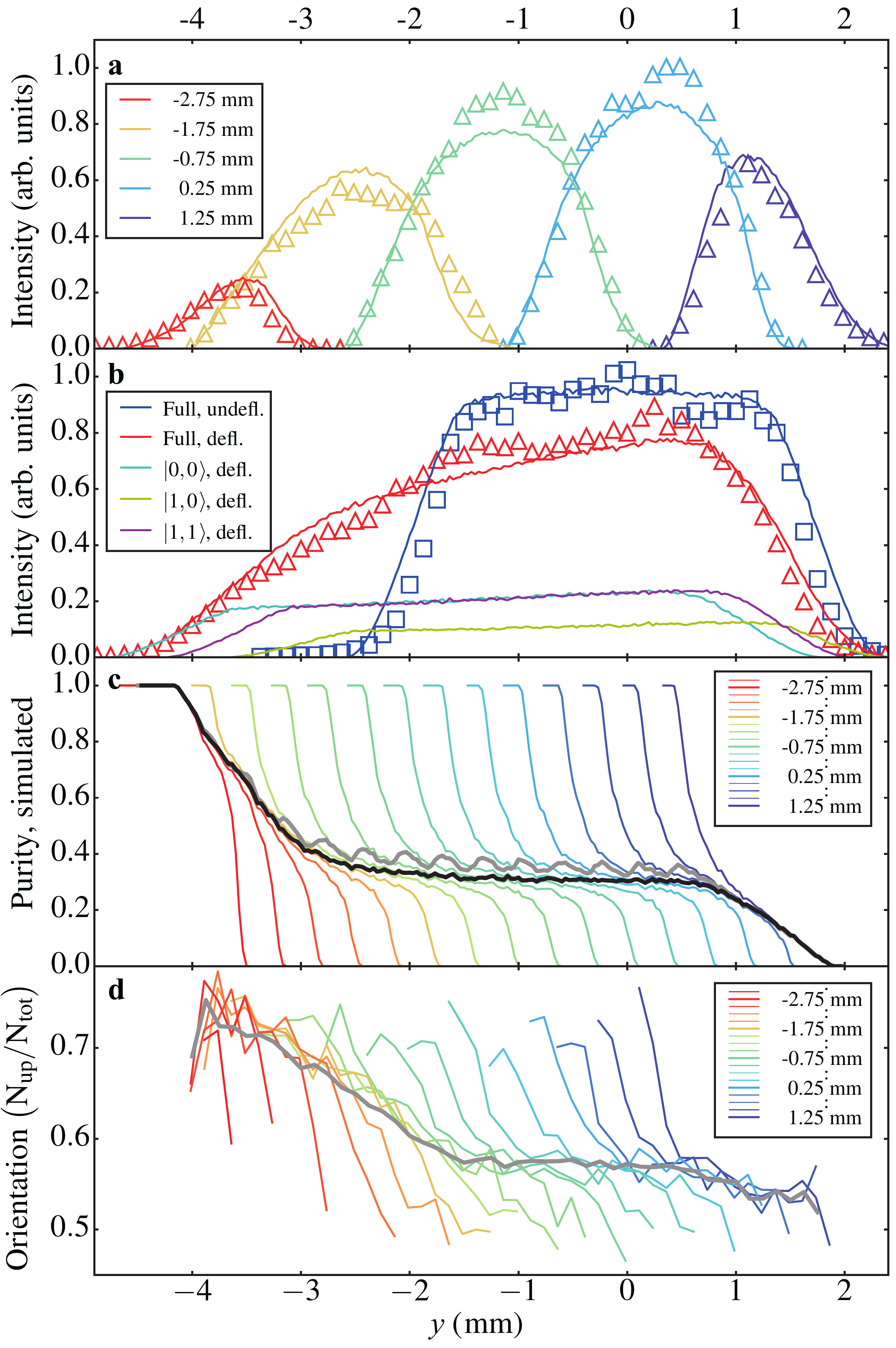}
   \caption{(Color online) (a) Experimental vertical density profiles for selected skimmer
     positions. (b) Experimental vertical density envelope profiles of the undeflected (blue
     squares) and deflected (red triangles) molecular beam. Simulated envelope profiles are shown as
     solid blue and red lines. Profiles of the individual rotational states \ket{0,0}, \ket{1,0} and
     \ket{1,1} are shown as cyan, yellow, and magenta lines, respectively. (c) Simulated purity of
     the \ket{0,0} state for various skimmer positions. (d) Experimental degree of orientation for
     various skimmer positions. See text for details.}
   \label{fig:deflection-profiles}
\end{figure}

\autoref{fig:deflection-profiles}~a shows the normalized measured (triangles) vertical density
profiles of the deflected molecular beam for selected skimmer positions at -2.75 mm (red), -1.75 mm
(yellow), -0.75 mm (green), 0.25 mm (cyan), and 1.25 mm (purple), respectively. A strong dependence
of the density profile on the position of the third skimmer is observed. Corresponding
simulated~\cite{Filsinger:JCP131:064309} beam-density profiles are shown as solid lines. Each
simulated profile fits the corresponding measured profile assuming a rotational temperature of
0.8~K.

\autoref{fig:deflection-profiles}~b shows the normalized measured vertical density profiles of the
undeflected (blue squares) and deflected (red triangles) molecular beam. The profiles are obtained
from the envelope of 18 measured single profiles recorded at specific skimmer positions between
-3.0~mm and 1.25~mm with a relative step size of 0.25~mm. All molecules are deflected downwards when
20\;kV and $-17.5$\;kV are applied to the deflector electrodes, as all quantum states are high-field
seeking at the electric field strenghts experienced inside the deflector. The corresponding
simulated beam-density profiles are shown as solid blue and red lines. The profiles of the
individual rotational states \ket{0,0}, \ket{1,0} and \ket{1,1}, which are the states that are
deflected the most, are shown as cyan, yellow, and magenta lines, respectively. Again, the simulated
profiles fit best to a rotational temperature of 0.8~K. For positions $y<-3.5$~mm practically only
the \ket{0,0} and \ket{1,1} states are populated. According to the simulations a pure ground state
population is observed for positions below $y=-4.2$~mm.

The simulated purity $p_i = N_{\ket{0,0}_i}/N_{i}$ of the \ket{0,0} state is shown for each specific
skimmer position $i$ as colored lines in \autoref{fig:deflection-profiles}~c. The number of
molecules in the absolute ground state and the total number of molecules is denoted by
$N_{\ket{0,0}_i}$, and $N_{i}$, respectively. The purity changes continuously from 0 to 1 with
decreasing laser focus positions for all skimmer positions. The simulated purity of the molecular
beam without the last skimmer is shown as a black line. A gray line represents the weighted mean of
the purity obtained for each position $p(y)=\sum p_i(y)N_i(y)/\sum N_i(y)$.

\autoref{fig:deflection-profiles}~d shows the measured degree of orientation
$o(y)=N_{\mathrm{up}}/N_{\mathrm{down}}$ across the deflected molecular beam for every skimmer
position $i$ as colored lines. As previously shown~\cite{Nielsen:PCCP13:18971, Kienitz:CPC17:3740},
the degree of orientation is a good measure of the purity. The degree of orientation increased for
decreasing vertical positions for each skimmer position $i$. The gray line shows the weighed mean of
the degree of orientation obtained from the individual skimmer positions
$o(y)=\sum o_i(y)N_i(y)/\sum N_i(y)$. An average degree of orientation of 0.58 was observed at
positions between $y=$-1.5 and $y=$1\,mm where all states are present. The degree of orientation
decreases to 0.54 for increasing $y$ with $y>1$\,mm.  In that region only higher states ($J>0$) are
present. The degree of orientation rises from 0.58 to 0.75 in the region between $y=-1.5$\,mm and
$y=-4$\,mm. Only states with $J<2$ are present in that region. The ground state purity and, therefore,
the degree of orientation is increased for decreasing vertical positions $y$.

\subsection{Simulated deflection of unpolar molecules}
\label{sec:deflection-unpolar}
\begin{figure}
   \includegraphics[width=\figwidth]{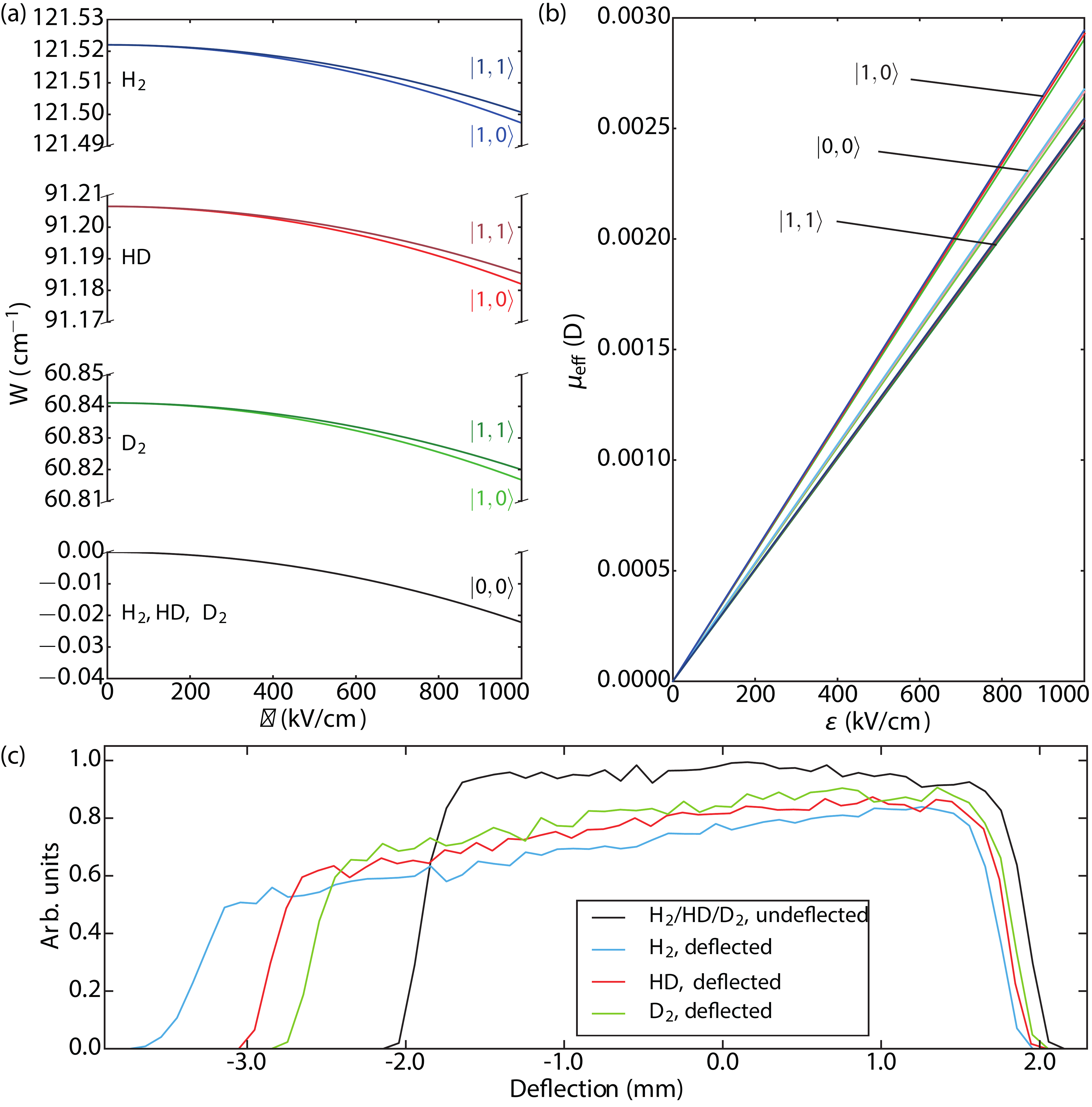}
   \caption{(Color online) (a) Stark energies and (b) effective dipole moments of the $J=0,1$ states
     of hydrogen molecule isotopologues. Both plots use the same color coding for the rotational
     states. (c) Vertical molecular beam profiles without deflection (black) and with deflection for
     H$_2$ (blue), HD (red), and D$_2$ (green).}
   \label{fig:deflection:hydrogen}
\end{figure}

To study further the performance of the new deflector we have computationally investigated its
applicability to the deflection and separation of non-polar neutral molecules. The Stark energy
curves and effective dipole moments of the $J=0,1$ states of the hydrogen isotopologues H$_2$, HD,
and D$_2$, are shown in \autoref{fig:deflection:hydrogen}~a and b.
In these calculatons rotational and centrifugal distortion-constants were taken from
references~\onlinecite{Orcutt:JCP39:605, Hamaguchi:MP43:1311} and polarizability anisotropies as
specified in \autoref{tab:polarizability}.
\begin{table*}
   \centering
   \begin{tabular}{llcllcc}
     \hline\hline
     molecule & mass & dipole moment & polarizability $\alpha_\parallel$
     & polarizability $\alpha_\perp$ & mean polarizability/mass & reference \\
     & (u) & (D) & ($10^{-24}$\,cm$^3$) & ($10^{-24}$\,cm$^3$) &
     ($10^{-24}$\,cm$^3$\,u$^{-1}$) & \\
     \hline
     OCS & 60 & 0.71 & 1.14 & 0.75 & 0.0126 & \onlinecite{Alms:JCP63:3321} \\[1ex]
     H$_2$ &  2 & 0 &  1.00220 & 0.70177 & 0.4510 & \onlinecite{Rychlewski:MP41:833} \\
     HD &  3 & 0 &  0.99483 & 0.69873 & 0.2987 & \onlinecite{Rychlewski:MP41:833} \\
     D$_2$ &  4 & 0 &  0.98556 & 0.69484 & 0.2222 & \onlinecite{Rychlewski:MP41:833} \\
     N$_2$  & 28 & 0 & 2.2146 & 1.5188 & 0.0625 & \onlinecite{Buldakov:JCP132:164304} \\
     CH$_4$ & 16 & 0 & 2.4442 & 2.4442 & 0.1528 & \onlinecite{Buldakov:JCP132:164304} \\
     C$_6$H$_6$ & 78 & 0 & 6.67 & 12.27 & 0.1094 & \onlinecite{Bridge:PRSA295:334} \\
     \hline\hline
   \end{tabular}
   \caption{Polarizabilities, dipole moments, and masses of some selected molecules discussed in
      this work.}
   \label{tab:polarizability}
\end{table*}
All Stark energies decrease as a function of electric field strength $\epsilon$ and,
correspondingly, all effective dipole moments \mueff are positive and, therefore, attracted to
regions of stronger electric field. A linear increase of the effective dipole moments \mueff as a
function of electric field strength $\epsilon$ was observed for all quantum states and
isotopologues. The largest differences in the slope of \mueff were observed between different
initial rotational states. Only small differences were found between the three isotopologues in a
given specific rotational quantum state. The simulated deflection profiles for a molecular beam with
a rotational temperature of $T_r=1$~K are shown in \autoref{fig:deflection:hydrogen}~c. The speed of
the simulated molecular beam was 385~m/s, which was demonstrated for molecular beams seeded in
krypton~\cite{Wohlfart:thesis:2008}. The black line depicts the overall profile for the undeflected
beam, without any field applied. It has an identical shape for all isotopologues. The colored lines
are the deflection profiles of the individual isotopologues for an applied maximum field strength of
363~kV/cm. All hydrogen isotopologues were deflected significantly in the simulation. The largest
deflection was observed for hydrogen which has the smallest mass of the three isotopologues.

\section{Discussion}
\label{sec:discussion}
The new HV-deflector is combining several advantages of the designs, that were used before. The
deflector used by Wrede and Stern in the 1920s is distinguished by its simplicity (a wire parallel
to a plain surface), but suffered in the resulting deflection force~\cite{Wrede:ZP44:261}. The
$a$-type deflector, as it was used by Chamberlain and Zorn~\cite{Chamberlain:PR129:677} increased
the force at a given electric potential. Nevertheless, it suffered from a deflection limit given by
the molecules crashing into the rod. The group of Markus Arndt developed a deflector with a
relatively large region of constant deflection force at the expense of the deflection
strength~\cite{Stefanov:MeasSciTech19:055801}. The deflector introduced here provides a strong
deflection force combined with a large region where molecules can be deflected, although the force
is not constant in this area. For a molecular beam with a small diameter, however, the force is in
good approximation constant. The maximum achieved field strength in the $b$-type deflector was
363~kV/cm. This is higher than the field strengths we typically achieve in a $a$-type deflector
which are in the order of 220~kV/cm. Moreover the highest field strength of the $a$-type deflector
is located far outside the region of the molecular beam (see~\autoref{fig:shape}). For a $b$-type
deflector the molecules are deflected into the region of highest field strength. The demonstrated
degree of deflection was sufficient to completely separate a part of the molecular beam from the
helium seed gas.

\begin{figure}
   \centering
   \includegraphics[width=\figwidth]{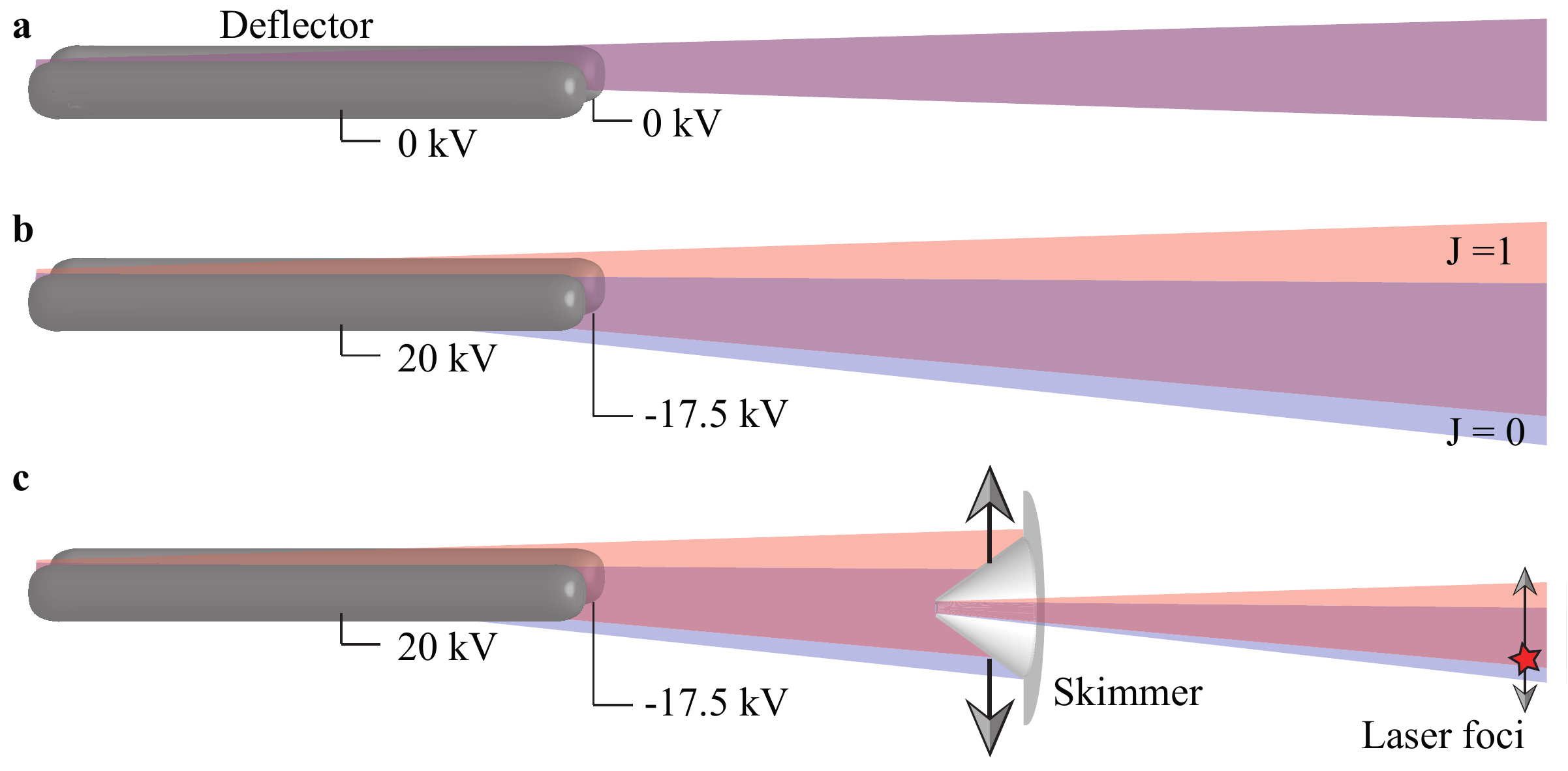}
   \caption{(Color online). Sketch of the dispersion of the molecular beam passing the deflector;
      the gray rods depict the deflector and the purple area the cross section of the molecular
      beam. (a) deflector switched off. (b) deflector switchen on. (c) influence of the skimmer.}
   \label{fig:results:skimmer}
\end{figure}

Both, the measured degree of orientation as well as the simulated purity show an increase towards
small laser positions for each specific skimmer position. This is attributed to the molecular beam
being dispersed according to the molecules' effective dipole moments or, correspondingly, according
the specific rotational states. The field dressed state that correlates adiabatically with the field
free rotational ground state of the molecule is most polar and therefore deflected the most. The
ground state gives rise to the highest degree of orientation. This is reflected by the increased
degree of orientation and purity in the deflected part of the molecular beam which accounts for the
larger contributions of low rotational quantum states. In addition, a decrease of the degree of
orientation in regions where the excited rotational states remain (right-hand part of the specific
molecular beam profiles) is observed. \autoref{fig:results:skimmer} sketches the dispersion of the
molecular beam passing the deflector in the cases of the deflector being switched off (a) and
switched on (b). The movable skimmer selects a part of the beam (c), which subsequently disperses
further due to the distinct transverse momenta of molecules in the different eigenstates. Therefore,
a quantum state separation is obtained for each specific skimmer position.

The weighted mean of the purity provides a good measure for the purity of the unskimmed molecular
beam. The region with the largest difference between the weighted mean purity and the unskimmed
purity is the central part of the curves where the weighted mean is increased compared to the purity
of the unskimmed molecular beam. This is attributed to the different shapes of the deflected ground
state profile and the deflected profile of the rotationally excited states. The difference between
the two profiles is small at the low temperatures of our molecular beam, and, therefore, the
weighted mean purity approximates the real purity. The oscillations for the weighted mean purity in
the central region is ascribed by the finite number of skimmer positions. We expect the weighted
mean to underestimate the degree of orientation without skimmer for the same reasons as discussed
before for simulations of the ground state purity. A direct comparison of the measured degree of
orientation and the simulated purity or quantum state distribution is, however, not possible due to
the highly non adiabatic orientation dynamics of the rotationally excited
states~\cite{Kienitz:CPC17:3740}. The character, and, therefore, the degree of orientation, of the
specific quantum states was changed in the mixed field due to passage through real and avoided
crossings while the laser pulse intensity was rising. Our simulations show that it should be
possible to obtain a pure ground state sample for each specific skimmer position in the most
deflected part of the molecular beam. Furthermore, from our simulations we saw that the column
density at this position is given by about 10\% of the peak column density. This is also reflected
in the measurements. The fluctuations in the degree of orientation in the most deflected part is
attributed to the low statistics of these measurements.

The $b$-type deflector has the potential to separate non-polar molecules. For the hydrogen
isotopologues we observed a linear increase of the effective dipole moment with increasing electric
field strength. This is attributed to the small influence of the electric field on the field free
rotational quantum states since all molecules in~\autoref{tab:polarizability} are still in the
perturbative regime for the simulated field strengths.

We observe in our simulation that the deflector can separate the different isotopologues, similar to
previous conformer-separation experiments~\cite{Filsinger:ACIE48:6900,
  Kierspel:CPL591:130}. However, the different rotational states of each isotopologue are not
separable by the current deflector and molecular beam diameter and divergence. Thus a nuclear-spin
purification, as previously demonstrated for water~\cite{Horke:ACIE53:11965}, is not possible with
the current setup. This is due to the fact that the Stark effect in the currently available dc
electric fields is dominated by the diagonal polarizability matrix element, \ie, essentially the
scalar polarizability, whereas a significant influence of the off-diagonal matrix elements would
render the states inequivalent. Our simulations show that a significant difference of effective
dipole moments and, therefore, a separability of the $J=0$ and $J=1$ states, would be obtained for a
25\,$\mathrm{\mu}$m wide non-divergent molecular beam at the current maximum field strength of 363~kV/cm.

Our simulations predict that spherical top molecules such as methane (CH$_4$) are
deflected fairly well, \eg, about 0.5~mm under the current experimental conditions. This might be
beneficial for experiments where the molecules have to be separated from the seed gas. However,
since for spherical-top molecules $\alpha_{\perp}=\alpha_{\parallel}$, their rotational quantum
states cannot separated due to their static polarizability at all.

\section{Summary}
\label{sec:summary}
In conclusion, we have shown that the deflection of cold molecular beams with an inhomogeneous
static electric field, given by the $b$-type-shaped deflector enables the selection and a strong
spatial separation of the most polar quantum states, i.e., the lowest-lying rotational states.
Furthermore, our simulations showed that the deflector will allow to separate isotopologues of non
polar molecules and that it enables background-gas free measurements of a whole new group of
molecules. In comparison with the $a$-type deflector it has the advantage that the highest field
gradient is located close to the molecular beam position which results in a stronger deflection. In
addition, due to the open structure of the deflector, strongly deflected molecules do not crash into
the electrodes. This open structure also enables its utilization in merged beam
experiments~\cite{Henson:Science338:234}. Furthermore, a stronger deflection could be reached by
optimizing the shape of the electrodes and the resulting field gradients. This strong deflection
might be used to reduce the length of the deflector to increase the total molecular density at the
region of interest.

The ability to disperse molecular beams by inhomogeneous electric fields is not limited to
rotational state selection of small linear molecules. The new design makes a better separation of
structural isomers of complex molecules as well as different sizes of individual clusters
possible~\cite{Chang:IRPC34:557}. These clean, well-defined, and spatially separated samples allow
for novel experiments such as femtosecond pump–probe measurements, gas-phase x-ray or electron
diffraction, or tomographic reconstructions of molecular orbitals. In addition it could be useful
for isolating molecular signals in high-harmonic generation and attosecond experiments. Furthermore,
state selection by deflection is highly advantageous in order to increase the degree of alignment
and orientation to study complex molecules in the molecular frame~\cite{Holmegaard:NatPhys6:428,
   Holmegaard:PRL102:023001}.

\section{Acknowledgements}
Besides DESY, this work has been supported by the \emph{Deutsche Forschungsgemeinschaft} (DFG)
through the excellence cluster ``The Hamburg Center for Ultrafast Imaging -- Structure, Dynamics and
Control of Matter at the Atomic Scale'' (CUI, EXC1074), the European Research Council through the
Consolidator Grant COMOTION (ERC-Küpper-614507), and the Helmholtz Association ``Initiative and
Networking Fund''.

\bibliography{string,cmi}
\end{document}